\documentclass{aa}
\usepackage{graphicx}

\begin{document}
  \title{Cerium: the lithium substitute in post-AGB stars\thanks{based on
   observations collected at the European Southern Observatory at Paranal in
   Chile (66.D-0171A) and at the Roque de los Muchachos Observatory at La
   Palma, Spain}}

\author{
Maarten Reyniers\thanks{Scientific researcher of the Fund for
Scientific Research, Flanders}$^{1}$,
Hans Van Winckel\thanks{Postdoctoral fellow of the Fund
for Scientific Research, Flanders}$^{1}$,
Emile Bi\'emont\thanks{Research Director of the FNRS}$^{2,3}$ \and
Pascal Quinet\thanks{Research Associate of the FNRS} $^{2,3}$}
\offprints{Maarten Reyniers, \email{maarten.reyniers@ster.kuleuven.ac.be}}
\institute{$^1$Departement Natuurkunde en Sterrenkunde, K.U.Leuven,
Celestijnenlaan 200B, B-3000 Leuven, Belgium\\
$^2$Astrophysique et Spectroscopie, Universit\'e de Mons-Hainaut, B-7000 Mons,
Belgium\\
$^3$Institut de Physique Nucl\'eaire, Atomique et de Spectroscopie (IPNAS),
Universit\'e de Li\`ege, B-4000 Li\`ege, Belgium}
\date{Received 1 August 2002 / Accepted 14 October 2002}
\titlerunning{Ce the Li substitute}
\authorrunning{M. Reyniers et al.}

%******************** ABSTRACT ****************************************
\abstract{In this letter we present an alternative identification for the
line detected in the spectra of s-process enriched low-mass post-AGB stars 
around 6708\,\AA\, and which was interpreted in the literature as due to Li. 
Newly released line lists
of lanthanide species reveal, however, the likely identification of the line
to be due to a Ce\,{\sc ii} transition. We argue that this identification is
consistent with the Ce abundance of all the objects discussed in the 
literature and conclude that in none of the low-mass s-process enriched post-AGB stars there 
is indication for Li-production. 
}
\maketitle

%******************** INTRODUCTION ************************************
\section{Introduction}
  Few chemical species are so important to guide our theoretical
understanding of both stellar and primordial nucleosynthesis as lithium (Li). 
The understanding of the solar Li abundance, the cosmic Li abundance
and the evolution of Li in our Galaxy is, 
however, complicated and at least three different production sites must 
be explored: (1) big-bang nucleosynthesis;
(2) spallation by cosmic ray particles on interstellar matter nuclei, and (3)
stellar nucleosynthesis, mainly during the AGB evolution
(e.g. Travaglio et al. 2001 and references therein).

  The production of Li in AGB stars is an important ingredient 
in the deciphering of the Galactic chemical evolution.
This production site is well documented observationally by the 
detection of several Li-rich AGB giants both in our 
Galaxy (e.g. Garc\'{\i}a-Lario et al. 1999) and in the Magellanic Clouds
(e.g. Smith \& Lambert 1989; Plez et al. 1993). Also the in situ
production of Li by the Cameron-Fowler transport
mechanism is rather well understood theoretically, by the process dubbed
`hot-bottom burning' in which the base of the convective stellar envelope
penetrates hot enough layers for nucleosynthesis to take place
and which guarantees quick transport of fresh $^{7}$Li to cooler 
layers to prevent destruction
by proton capture (e.g. Sackmann \& Boothroyd 1992; 
Lattanzio \& Forestini 1999). This hot-bottom process is, however, 
only expected to be active in intermediate mass stars of 
around 4-5 M$_{\odot}$ depending on the metallicity 
(e.g. Travaglio et al. 2001).

  The high Li abundance of low-mass post-AGB stars described recently
in the literature came, therefore, as a surprise. Indeed, several objects
with a low metallicity in the post-AGB phase of stellar evolution were
described to be enhanced in Li to such extent that production of Li
has to be invoked to explain the derived abundances. The objects
are IRAS\,05341+0852 (Reddy et al. 1997), IRAS\,22272+5435
(Za\v{c}s et al. 1995; Reddy et al. 2002), IRAS\,Z02229+6208 and
IRAS\,07430+1115 (Reddy et al. 1999), IRAS\,05113+1347 (Reddy et al. 2002)
and they all combine a high Li abundance with a strong overabundance of neutron
capture elements and a C/O number ratio larger than one. Their effective
temperature and the presence of a significant amount of circumstellar dust
indicate these objects to be in a post-AGB evolutionary stage. Their low
metallicity and kinematics show these objects to be of low initial mass
and for these stars, the 3rd dredge-up phenomenon, although not well
understood by itself, is not expected to bring freshly produced Li to the
stellar photosphere. 

   The presence of Li in these stars was inferred from the famous Li
resonance doublet at 6707.76\,\AA\ and 6707.91\,\AA\ but in all stars, the
doublet is shifted redwards by about 0.2\,\AA\ compared to the expected
position based on the other photospheric lines. A 
differential velocity of the Li line-forming region is invoked to
explain this shift.
An additional indication for this identification was that similar shifts
were found in components of the Na D and K I alkali line profiles (e.g.
Reddy et al. 2002). Those profiles are, however, complex and 
often disturbed by emission components (see e.g. Figs. 5 and 6 of the
same paper), making the presence of such a component at the correct
velocity not clear. Contrary to the abovementionned older papers, 
the two newest quantitative Li abundances were
derived by spectrum synthesis modelling in the wing of the ''shifted Li line'' 
at the photospheric position (Reddy et al. 2002).

  Motivated by this unexplained velocity shift in the Li line, we investigated 
an alternative identification of this line which was also detected
in our high signal-to-noise (S/N) UVES spectra in several objects. Given
the facts that the stars are strongly enriched in s-process elements and 
that the line spectra of (ionised) s-process elements is not well documented,
we scanned the newest s-process line lists for possible identifications.
Here we report on our alternative identification of the 
line at 6708.1\,\AA\ as due to Ce\,{\sc ii}. The paper is organised as follows:
In Sect.~2 we document our newest high S/N-spectra
of some post-AGB stars; the next
section is devoted to the line list update of the lanthanides 
while Sect.~4 and 5 are devoted to our spectral analysis of the Li line 
region. We end this paper with a discussion in Sect.~6.

%******************** High S/N spectra with VLT+UVES ***********
\section{High S/N post-AGB spectra with VLT+UVES}
In the framework of our ongoing program to study the photospheric chemical
composition of stars in their last stages of evolution, high resolution,
high S/N spectra of a selected sample of 11 post-AGB objects were
obtained with the UVES spectrograph mounted on the 8.2\,m VLT-KUEYEN telescope.
The observations were carried out in service mode during two periods
(04-07/2000 and 01-02/2001). The resolving power of these spectra 
varies between $\sim$55,000 and $\sim$60,000. Sample spectra can be found
in Fig.~\ref{fig:idnt} (the two spectra on top).

The chemical diversity present in the whole post-AGB class of stellar
objects (e.g. Van Winckel \& Reyniers 2001) is also reflected in 
this new sample: only three out of the eleven objects are clearly 
s-process enhanced, two of which are observed for the first time at high resolution.
We already completed a detailed abundance analysis for each of these objects,
but these will be published elsewhere.

%******************** The Ce II D.R.E.A.M. line data ******************
\section{The Ce\,{\sc ii} D.R.E.A.M. line data}

One of the problems in dealing with strongly s-process enriched objects is
the lack of accurate atomic data of many s-process elements. Besides the more
general lack of accurate oscillator strengths, even the wavelengths of the 
transitions of neutral and ionised s-process species are badly known. 
Given the fact that the optical spectrum of the strongly enriched objects like 
IRAS\,05341+0852 is completely dominated by such transitions (Fig.~\ref{fig:idnt}), this 
limits severely the astrophysical interpretation of such data.

The main purpose of the ``Database on Rare-Earths At Mons University'' (D.R.E.A.M.),
created recently and accessible at the address 
http://www.umh.ac.be/$\sim$astro/dream.shtml,
is to provide such an update concerning the radiative
properties of rare-earth atoms and ions. These properties are obtained by systematic 
and detailed calculations performed within the framework of a pseudo-relativistic 
Hartree-Fock (HFR)
method (Cowan 1981) including core-polarization effects (see e.g. Quinet et al. 1999).
In the particular case of doubly ionized cerium, 
about 15000 lines with wavelengths based on {\sl experimental} energy levels 
are listed in the database for which atomic data were obtained 
using the computational procedures described
by Palmeri et al. (2000) and Zhang et al. (2001). 
This linelist extends considerably
the one reported in Kurucz' database which only contains the Ce\,{\sc ii} lines observed by
Meggers et al. (1975).

%******************** The identification of the 6708.1 line ***********
\section{The identification of the 6708.1\,\AA\ line}

\begin{figure}\caption{{\em upper panel}: Line identification of the 6708\,\AA\ region for
three heavily s-process enriched objects. The Ce\,{\sc ii} 6708.099\,\AA\ line 
matches exactly the ``shifted Li line''. Note also that for some lines in
these objects a definitive identification is still missing.
{\em lower panel}: Spectrum synthesis (full line) of the 6708\,\AA\
region for IRAS\,05341+0852 (dots).
For this synthesis we used the abundances found in our abundance analysis
without any modification, in order to get an idea of the quality of the
line list in this region. The lithium doublet was {\em not} included in our
list.}\label{fig:idnt}
\resizebox{\hsize}{!}{\includegraphics{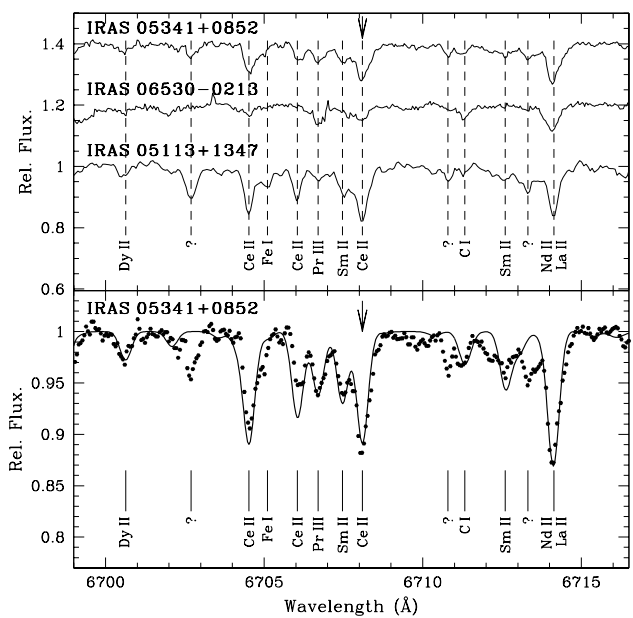}}
\end{figure}

As soon as the Ce\,{\sc ii} line list was released, we realized that the Ce\,{\sc ii}
line at 6708.099\,\AA\ ($\chi$\,=\,0.701\,eV, $\log\,gf$\,=\,$-$2.12) was an obvious candidate 
to identify the ``shifted Li line'' in the s-process enriched post-AGB stars. 
We decided to make a complete identification of the
6708\,\AA\ region for the s-process enriched post-AGB stars by extracting all
spectral lines from the VALD database (Kupka et al. 1999) from 6695
to 6725\,\AA. From this list, we removed the lithium doublet. Then, we added
the Pr\,{\sc iii} lines and the Ce\,{\sc ii} 6708.099\,\AA\ line from the D.R.E.A.M.
database. Although the accuracy of an oscillator strength is always hard to assess, it is
 reasonable to consider that the f value of the 6708.099\,\AA\ transition is probably accurate 
within a few percent (typically 10 to 15\%) in view of the moderate complexity of the
 electronic configurations involved in Ce\,{\sc ii}.
We chose not to replace all other Ce\,{\sc ii} VALD $\log\,(gf)$ values by  D.R.E.A.M. 
data in order not to compromise a comparison with previous results (see later).
For two carbon lines in our list, we modified the log$gf$ values to
more recent values (Hibbert et al. 1991).

We estimated the equivalent widths for all the lines in this region using
abundances determined from our new spectra. These calculations were made 
using R.L. Kurucz model atmospheres in combination with C. Sneden's latest version (April 2002)
of his LTE line analysis program MOOG. We could identify in this way 
most lines in the Li-region (see Fig.~\ref{fig:idnt}),
including the line previously identified as the ``shifted Li line'',
indicated with an arrow.

Moreover, we synthesised the 6708\,\AA\ region for IRAS\,05341+0852
(Fig.~\ref{fig:idnt}). The aim of this synthesis was not to derive abundances, 
but to test the quality of our constructed line list. It turned out that most of the lines were
adequately fitted, but some lines are clearly still missing. It is striking that even
relatively strong lines, such as the line at $\sim$6702.7\,\AA, are not yet
identified, but are very probably also of s-process origin. 
The region close to the 6708.1\,\AA\ line is
very well fitted with the new Ce\,{\sc ii} covering the ``shifted Li line''.

%******************** The ultimate lithium death-blow *****************
% \section{The ultimate lithium death-blow}

\begin{table*}\caption{Comparison between the Ce abundance calculated from the
6708.099\,\AA\ line and the calculated or published Ce abundance for the
post-AGB stars showing the ``lithium-shift''. Most stars show a good to
excellent agreement between the two abundances, implying a safe identification
of the 6708.1\,\AA\ line as being due to Ce\,{\sc ii}.}\label{tab:abnd}
\begin{center}
\begin{tabular}{rccccc}
\hline
\multicolumn{1}{c}{Object} & Ce\,{\sc ii} 6708 & Ce abund.   & calculated\,or\,published    & adopted model & ref.\\
       & equivalent    & from 6708 & Ce abundance & T$_{\rm eff}$, $\log g$, $\xi_t$, [M/H] & \\
       & width (m\AA)  &  line  ($\log\epsilon$)                    & $\log\epsilon$ ($n$, $\sigma$)& 
                                                                     (K, cm\,s$^{-2}$, km\,s$^{-1}$)     \\
\hline
IRAS\,05341+0852 & 53.5 & 3.40 & 3.35 (23, 0.15) & 6500, 1.0,\phantom{5}3.5,\phantom{5}$-$1.0 & 1\\ 
IRAS\,06530$-$0213 & 21.5 & 3.54 & 3.34 (18, 0.14) & 7250, 1.0,\phantom{0}5.0,\phantom{5}$-$0.5 & 1\\
\hline
IRAS\,05113+1347 & 106.\phantom{7} & 2.83 & 2.84 (12, 0.20) & 5250, 0.25,4.5,\phantom{5}$-$0.5 & 2\\
IRAS\,22272+5435 &  85.\phantom{3} & 2.99 & 2.76 \phantom{1}(6, 0.14) & 5750, 0.5,\phantom{5}4.5,\phantom{5}$-$1.0 & 2\\
\hline
IRAS\,Z02229+6208 & 65.\phantom{0} & 2.75 & 2.09 \phantom{1}(4, 0.04) & 5500, 0.5,\phantom{5}4.25,$-$0.5 & 3 \\
IRAS\,07430+1115 & 40.\phantom{0} & 2.90 & 2.45 \phantom{1}(4, 0.12) & 6000, 1.0,\phantom{5}3.5,\phantom{5}$-$0.5 & 3\\
\hline
\multicolumn{6}{l}{References for columns 2, 4 and 5: (1) this letter, (2) Reddy et al. 2002, (3) Reddy et al. 1999}\\
\end{tabular}
\end{center}
\vskip -.156 cm
\end{table*}

As a final test of our lithium substitute, we calculated the Ce abundance from
the 6708.099\,\AA\ line for all post-AGB stars showing the ``shifted
Li line'', and compared it to the calculated or published Ce abundance.
The equivalent widths for the different stars are obtained as follows:
\begin{itemize}
\vspace{-.24 cm}
\item IRAS\,05341+0852 and IRAS\,06530$-$0213: measured from
KUEYEN+UVES spectra. The abundance analyses of these objects will be published
elsewhere.
\item IRAS\,05113+1347 and IRAS\,22272+5435: measured from Fig.~7 in Reddy et al. (2002).
\item IRAS\,Z02229+6208 and IRAS\,07430+1115: taken from Tab. 10 in Reddy et al. (1999).
\end{itemize}
\vspace{-.24 cm}
For IRAS\,05113+1347 we have also an own WHT+UES spectrum (see Fig.
\ref{fig:idnt}) from which we obtained
$W_{\lambda}(6708.1{\rm\AA})$\,=\,81\,m\AA.
However, we chose to use the Reddy et al. $W_{\lambda}$ to be consistent with
their abundance analysis and model for this star. For IRAS\,22272+5435,
Za\v{c}s et al. (1995) mention $W_{\lambda}(6708.1{\rm\AA})$\,=\,106\,m\AA.

From the comparison summarized in Tab. \ref{tab:abnd}, we can conclude that
there is a good to excellent
agreement between the Ce abundance calculated from the 6708.099\,\AA\
line and the calculated or published Ce abundance. Only for the stars
IRAS\,Z02229+6208 and IRAS\,07430+1115 the difference in abundance is
larger (but still $<$0.7\,dex). However, the Ce abundance for these stars
is based on only four lines and therefore it should be reanalysed. It is clear,
though, that also for these stars the identification of the
6708.1\,\AA\ line as a ``shifted Li line'' is much less probable
than as a Ce\,{\sc ii} line. Moreover, it is unlikely that the same velocity shift
should prevail for circumstellar Li or Li in photospheric downdrafts 
(e.g. Reddy et al. 2002) in all post-AGB objects.

%******************** Lithium upper limits ****************************
\section{Lithium upper limits}

As seen in Fig. \ref{fig:idnt}, the new Ce line matches the ``shifted Li line''
very well. Reddy et al. (2002) computed a spectrum synthesis
of this region and derived the Li abundance from an indication of extra absorption
at the photospheric wavelength, without modelling the ''shifted Li line'' itself (see Fig. 7 of
their paper). Other Li abundances covered in the discussion and Tab. 8 of the same paper 
were derived from the shifted line equivalent width.

\begin{figure}
\caption{Fine synthesis of the 6708\,\AA\ region. Observed spectra (dots) are
compared with spectrum syntheses (solid lines) with three different Li
abundances: (1) no lithium, (2) our upper limit and (3) the Li abundance
reported in literature. The hypothetical Ce\,{\sc ii} line at
6707.74\,\AA\ from Lambert et al. (1993) was also included in the line
list and marked on the figure in italic.}\label{fig:samplsnth}
\resizebox{\hsize}{!}{\includegraphics[angle=-90]{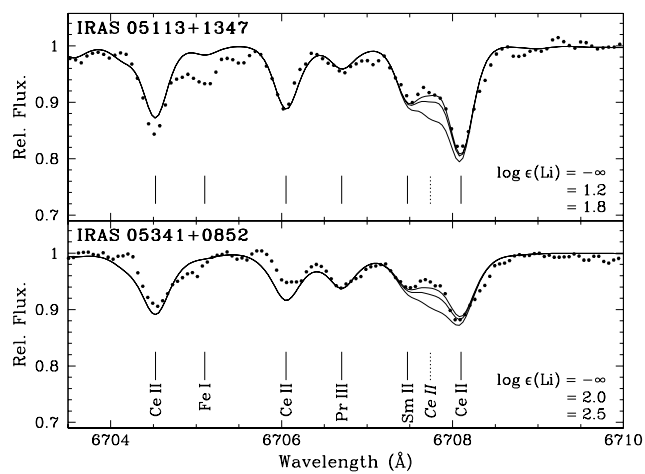}}
\end{figure}

To test this result, we derived
upper limits for the lithium abundance in the three stars from
Tab.~\ref{tab:abnd} of which we have spectra, one of which was discussed and shown
in Fig. 7 of Reddy et al. (2002).  For this purpose, we used the same atomic and
molecular line list as Reddy et al. (2002) including the s-process line 
at 6707.74\,\AA\ introduced by Lambert et al. (1993) who assumed it to be a
Ce\,{\sc ii} transition.
Sample syntheses and upper limits are given in Fig.~\ref{fig:samplsnth}.
If the suspected s-process lines at 6707.74\,\AA\ is excluded, the upper limits
shift up by $\sim$0.5\,dex for the cooler IRAS\,05113+1347 and $\sim$0.2\,dex 
for IRAS\,05341+0852. The upper limit for IRAS\,06530-0213 (not shown) was
$\log \epsilon$(Li) $<$ 2.9.
It is clear that no detection of a contribution due to Li could be made when
taken all lines of this complex region into account.

%******************** Conclusion **************************************
\section{Conclusion}
The line at 6708.1\,\AA\, which is seen in some heavily s-process enriched
post-AGB objects was identified in the literature as due to Li, despite the
redward shift of about 0.2\,\AA\ compared to other photospheric lines. 
The line was assumed to be formed in infalling circumstellar gas or photospheric
downdrafts to explain the shift. Using the recent atomic data on lanthanides released in
the D.R.E.A.M. project, we can safely identify this line as not being due to Li but
coming from a  Ce\,{\sc ii} line transition at 6708.099\,\AA.
Indeed, the Ce abundance derived from this single line is comparable
to the Ce abundance derived from other Ce\,{\sc ii} lines in the spectra. 
Moreover no inclusion of a significant amount of lithium at photospheric 
wavelength was needed in a synthesis of three objects, among which the most
s-process enriched object known. We claim that in none of the 
post-AGB objects there is evidence for in situ Li production and conclude that
there is no need to invoke special non-standard mixing during the AGB evolution to 
explain the claimed high abundances of this fragile element in these stars.
Note that this new line identification is no alternative for (super) Li rich lower luminosity
objects.

This result dramatically illustrates 
the importance of reliable and complete line lists when doing analyses of 
these strongly s-process 
enriched objects, the spectra of which are dominated by lines coming from
s-process elements.

%******************** ACKNOWLEDGEMENTS ********************************
\begin{acknowledgements}
MR and HVW, EB and PQ acknowledge support from the Fund for Scientific
Research -- Flanders and the FNRS (Belgium), respectively.
This research has made use of the Vienna Atomic Line Database (VALD2).
It is a pleasure to thank Prof. C. Sneden for his
continuous support of his MOOG Stellar Line Analysis Program.
\end{acknowledgements}

%******************** BIBLIOGRAPHY ************************************

\end{document}